\documentclass[preprint,amsmath,amssymb]{revtex4}

\usepackage{amsmath}
\usepackage{amssymb}
\usepackage{graphicx}
\usepackage{amsfonts}
\usepackage{dcolumn}
\usepackage{bm}

\newcommand \la{\langle}
\newcommand \ra{\rangle}
\newcommand \f{\frac}

\begin{document}

\title{Spatiotemporal dynamics on small-world neuronal networks: The roles of two types of time-delayed coupling}

\author{Hao Wu}
\author{Huijun Jiang}
\author{Zhonghuai Hou} \email{hzhlj@ustc.edu.cn}

\affiliation{Hefei National Laboratory for Physical Sciences at
 Microscales \& Department of Chemical Physics, University of
 Science and Technology of China, Hefei, Anhui 230026, China}

\date{\today}

\begin{abstract}
We investigate temporal coherence and spatial synchronization on small-world networks consisting of noisy Terman-Wang (TW) excitable neurons in dependence on two types of time-delayed coupling: $\{x_j(t-\tau )-x_i (t)\}$ and $\{x_j(t-\tau )-x_i(t-\tau)\}$. For the former case, we show that time delay in the coupling can dramatically enhance temporal coherence and spatial synchrony of the noise-induced spike trains. In addition, if the delay time $\tau$ is tuned to nearly match the intrinsic spike period of the neuronal network, the system dynamics reaches a most ordered state, which is both periodic in time and nearly synchronized in space, demonstrating an interesting resonance phenomenon with delay. For the latter case, however, we can not achieve a similar spatiotemporal ordered state, but the neuronal dynamics exhibits interesting synchronization transition with time delay from zigzag fronts of excitations to dynamic clustering anti-phase synchronization (APS), and further to clustered chimera states which have spatially distributed anti-phase coherence separated by incoherence. Furthermore, we also show how these findings are influenced by the change of the noise intensity and the rewiring probability. Finally, qualitative analysis is given to illustrate the numerical results.
\end{abstract}


\maketitle

\section{Introduction\label{Section1}}

Neuronal networks from living biological entities to various theoretical models have gained great research attention in recent years \cite{RMP06001213}. As we know, neuronal networks consist of chemically coupled or functionally associated neurons, the connections among them can be formed by electric synapse or chemical synapse. In the vertebrate cortex, a neuron can be connected to as many as $10^{4}$ postsynaptic neurons, so the way in which neurons process and transmit information among each other is an important subject of research. Many experimental evidences demonstrate that neurons transmit information by processing them into action-potential sequences (spike trains), and spatial synchronization as well as temporal coherence of neuronal spike trains are crucial for coding and transmission of information across the neuronal networks \cite{Rie1996, Ger2002}. In the past two decades, extensive research has been performed with the aim of analyzing spatial synchronization and temporal coherence of neuronal dynamics, and many insightful findings have been reported.

On one hand, neurons are noisy elements, where noise arises from both external (e.g., synapses) and internal (e.g., channels) sources. The effects of noise on firing dynamics of neuronal networks, especially synchronization and temporal coherence, have been widely studied. For instance, Gaussian white noise can induce coherence resonance in FitzHugh-Nagumo (FHN) and  Hodgkin-Huxley (HH) neuronal models \cite{PRL97000775, CPC04001602, CHA08023102}, Perc \emph{et al.} \cite{ EPL09040008, EPJ09000964} showed that channel noise in coupled HH neurons can control the spontaneous spike regularity, and the effects of correlated noise on spike coherence and spike firing rate of coupled neurons have also been investigated thoroughly in various neuronal systems \cite{PRL01098101, PHA08006679, CHA10033116} by Kurths. On the other hand, due to the complex topological connections, real neuronal networks often exhibit small-world and scale-free features \cite{NTR98000440, PRL05018102}, so spatiotemporal dynamics on complex neuronal networks has attracted increasing attention \cite{PRL00002758, PRL06238103, PRL07108101, CSF07000280}. In previous works, we found that spatial synchronization and temporal coherence, which are practically absent in regular networks, can be greatly enhanced by random shortcuts between the neurons \cite{CPC05001402, CPC06000579, PRE06046137}. Stochastic resonance on small-world neuronal networks via a pacemaker was also investigated by Perc \cite{PRE07066203, PLA09000964}, and Kwon \cite{PLA02000319} reported that coherence resonance can be considerably improved by small-world connectivity in networks of HH neurons. Moreover, spatial synchronization and stochastic resonance have also been extensively studied on scale-free networks \cite{PRL06164102, PRE04056207, PRE08036105, BPC09000175}.

As is well known, information transmission delays are inherent to the real neuronal networks because of the finite speed at which action potentials propagate across neuron axons and the time lapses occurring in both dendritic and synaptic processing. For example, the speed of signal conduction through unmyelinated axonal fibers is on the order of 1 m/s, resulting in time delays up to 80 ms for propagation through the cortical networks \cite{Kan1991}. It is thus important to understand how the dynamics of coupled neuronal ensembles are influenced by such delays. A number of interesting effects of time-delayed coupling on the qualitative and quantitative properties of neuronal dynamics have been reported in literature, including both chemical synapse coupling \cite{PRL95001570, PRE01021908, PRE05061904, PRE08036211, CSF10000096, CSF09000918, IJB08001189} and electric synapse coupling \cite{PRL04074104, PRE03056219, CPL05000543, PRL05238103, CHA10043140, PHA10003299, EPL08050008, PRE10037201, PRE09026206, CSF04000267, PLO11015851, BPC10000126, CHA08037108, PLA08005681, IJM10001201,CHA09023112,CPB10040548}. For instance, time delays through chemical synapse can induce synchronization in coupled integrate-and-fire \cite{PRL95001570} and bursting Hindmarsh-Rose (HR) neurons \cite{PRE08036211}, tame chaos on scale-free neuronal networks \cite{CSF10000096}. Moreover, Wang \emph{et al.} found synchronization transitions from chaotic to periodic motions in two coupled FHN neurons \cite{CSF09000918}, as well as transitions between in-phase and anti-phase synchronization in two coupled fast-spiking neurons \cite{IJB08001189}. Compared with chemical synapse, time delays through electric synapse are more usual in academic research. People have found that time delay through electric synapse can facilitate and enhance neuronal synchronization \cite{PRL04074104, PRE03056219, CPL05000543}, induce various spatiotemporal patterns \cite{PRL05238103}, enhance spatiotemporal order in coupled noisy small-world neuronal networks \cite{CHA10043140}. In addition, Perc and his cooperators have contributed some remarkable findings in this field, they found that information transmission delay can induce transition from zigzag fronts to clustering anti-phase synchronization \cite{PHA10003299} and further to regular in-phase synchronization on small-world neuronal networks \cite{EPL08050008}, intermittently induce synchronization transitions on scale-free neuronal networks \cite{PRE09026206, PLO11015851}. Furthermore, they also showed that delay can enhance coherence of spatial dynamics in small-world networks of HH neurons \cite{PLA08005681, IJM10001201} and induce multiple stochastic resonances on scale-free neuronal networks \cite{CHA09023112, CPB10040548}, they proved that delay-induced multiple stochastic resonances are robust to the changing of the scale-free networks, even when the nodes of the network are more than $10000$.

It is worth noting that in the aforementioned literature electric synapse coupling with delay is described by $\{x_j(t-\tau )-x_i (t)\}$, whereas as we know, there exists another important scheme of delays through electric synapse which is defined by $\{x_j(t-\tau )-x_i(t-\tau)\}$. This type of coupling has been widely used to investigate synchronization problems in various fields such as electric circuit \cite{IJB01001707, CSF05001285}, coupled pendulums \cite{PRE08056213}, delayed neural networks (DNNS) \cite{PLA06000318}, and general models \cite{PRL10068701, PRL10208701, PHA04000263, PHA06000024, PHA08002111, PLA06000263, PLA08007133}. In a recent paper, we have showed that the former type of delayed coupling can enhance spatiotemporal order in coupled neuronal systems \cite{CHA10043140}. However, little attention has been paid to the impact of the latter type of coupling scheme on the spatiotemporal dynamics of neuronal networks. Furthermore, to this day, the differences between the effects of these two types of delayed coupling on spatiotemporal dynamics in coupled neuronal systems have not been studied. Thus in this paper, we aim to extend the scope of above-mentioned investigations by comparing such differences. To do this, we investigate temporal coherence and spatial synchronization on small-world networks consisting of noisy Terman-Wang (TW) excitable neurons in dependence on these two types of time-delayed coupling. We show that with same coupling strength and delay time some distinct results induced by these two different types of coupling can be achieved. For the former case, it is found that time delays can dramatically enhance neuronal synchrony and temporal coherence. In addition, if the delay time is tuned to nearly match the intrinsic spike period of the neuronal network, the system dynamics reaches a most spatiotemporal ordered state, demonstrating an interesting type of resonance phenomenon with delay. For the latter case, however, we can never find a similar spatiotemporal ordered state, whereas as the delay time is increased, the neurons exhibit synchronization transition from zigzag fronts of excitations to dynamic clustering anti-phase synchronization, and further to clustered chimera states. Furthermore, we also show how these findings are influenced by the change of the noise intensity and the rewiring probability. Finally, we give some qualitative analysis to illustrate the numerical results.

The remainder of this paper is structured as follows. In Sec.\ref{Section2} Terman-Wang neuronal networks with two types of time-delayed coupling are employed to simulate neuronal dynamics. Main results are presented in Sec.\ref{Section3}, followed by conclusions and discussions in Sec.\ref{Section4}.

\section{Model Description\label{Section2}}

The two-variable Terman-Wang (TW) model is similar to the famous FHN model, it was first proposed by Terman and Wang \cite{PHD95000148} to simulate neuronal oscillations discovered experimentally in visual cortex. We consider here $N$ coupled TW neurons with delayed coupling, subjected to additive Gaussian white noises and external signal. The  system dynamics can be described by the following equations:
\addtocounter{equation}{1}
\begin{align}
\dot{x_{i}} &=3x_{i}-x_{i}^{3}+\alpha-y_{i}+I+D\xi_{i}(t)+G_{i} \tag{\theequation a} \label{Model-1a} \\
\dot{y_{i}} &=\psi[\gamma(1+\tanh(x_{i}/\beta))-y_{i}] \tag{\theequation b} \label{Model-1b}
\end{align}

\noindent Here, $i=1,2,\ldots,N$ specifies the neuron index, variables $x_i$ and $y_i$ denote the action potential and the channel activation level of neuron $i$, respectively. $x$ is a fast variable and $y$ is a
slow one. $\psi$ is a small parameter which measures the time scale separation between the dynamics of $x$ and $y$. $\alpha, \beta, \gamma$ are model specific parameters. We consider that all the neurons are
identical and fix $\psi=0.02,\alpha=1.99,\beta=0.1,\gamma=6.0$ throughout this paper unless specified otherwise. $I$ represents a homogeneous subthreshold periodic stimulus current delivered externally to the neurons, and we set $I=0.01\sin(2\pi t/9)$. It should be noted that changing the amplitude and frequency of such a subthreshold periodic stimulus $I$ does not affect main results of this work. $\xi_{i}(t)$ stands for independent Gaussian white noise with unit variance, i.e., $\la \xi_{i}(t)\ra=0 $, $\la \xi_i(t)\xi_j(t') \ra = \delta_{ij}\delta(t-t')$. For these parameters, an isolated TW neuron will stay in the rest state in the absence of noise. $G_{i}$ is the coupling term, which represents interaction between all other neurons and neuron $i$. In this paper, we employ two types of electric synapse coupling with delay, $\{x_j(t-\tau )-x_i (t)\}$ and $\{x_j(t-\tau )-x_i(t-\tau)\}$, thereinto, $\tau$ is the transmission delay. Thus the coupling term $G_{i}$ can be described by $\epsilon \sum\limits_{j}A_{ij}[x_{j}(t-\tau)-x_{i}(t)]$ and $\epsilon \sum\limits_{j}A_{ij}[x_{j}(t-\tau)-x_{i}(t-\tau)]$, where $\epsilon$ is the coupling strength. The adjacency matrix $\mathbf{A}$ denotes connectivity of the neuronal network, with entry $A_{ij}=A_{ji}=1$ if neuron $i$ and $j$ are connected, and $0$ otherwise, $A_{ii}$ is set to 0. Numerical integration of Eq.(1) is carried out by using explicit Euler method with time step 0.003. Finally, we should note\underline{} that for convenience, the two types of delayed coupling $\{x_j(t-\tau )-x_i (t)\}$ and $\{x_j(t-\tau )-x_i(t-\tau)\}$ are defined as type \uppercase \expandafter {\romannumeral 1} and type \uppercase \expandafter {\romannumeral 2} coupling respectively in the remainder of the paper.

\section{Results\label{Section3}}

\begin{figure}
\centerline{\includegraphics*[width=1.0\columnwidth]{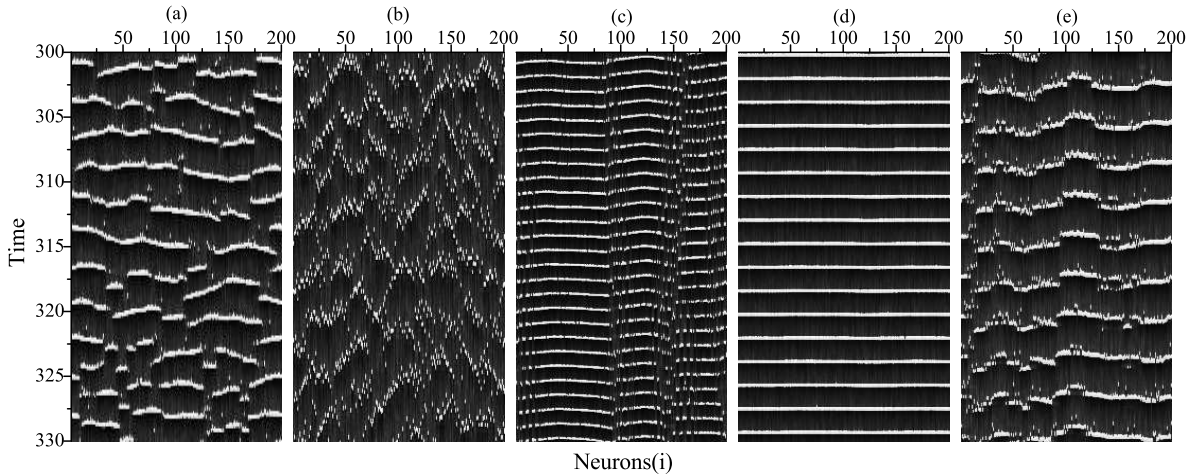}}
\caption{ Space-time plots of TW neurons with type \uppercase \expandafter {\romannumeral 1} coupling for different delay time $\tau$. From left to right, $\tau$ equals to $0, 0.3, 1.0, 1.8$ and $3.0$, respectively. Other parameter values are $N=200$, $K=8$, and $D=0.6$.
\label{Pattern-type1}}
\end{figure}

\begin{figure}
\centerline{\includegraphics*[width=1.0\columnwidth]{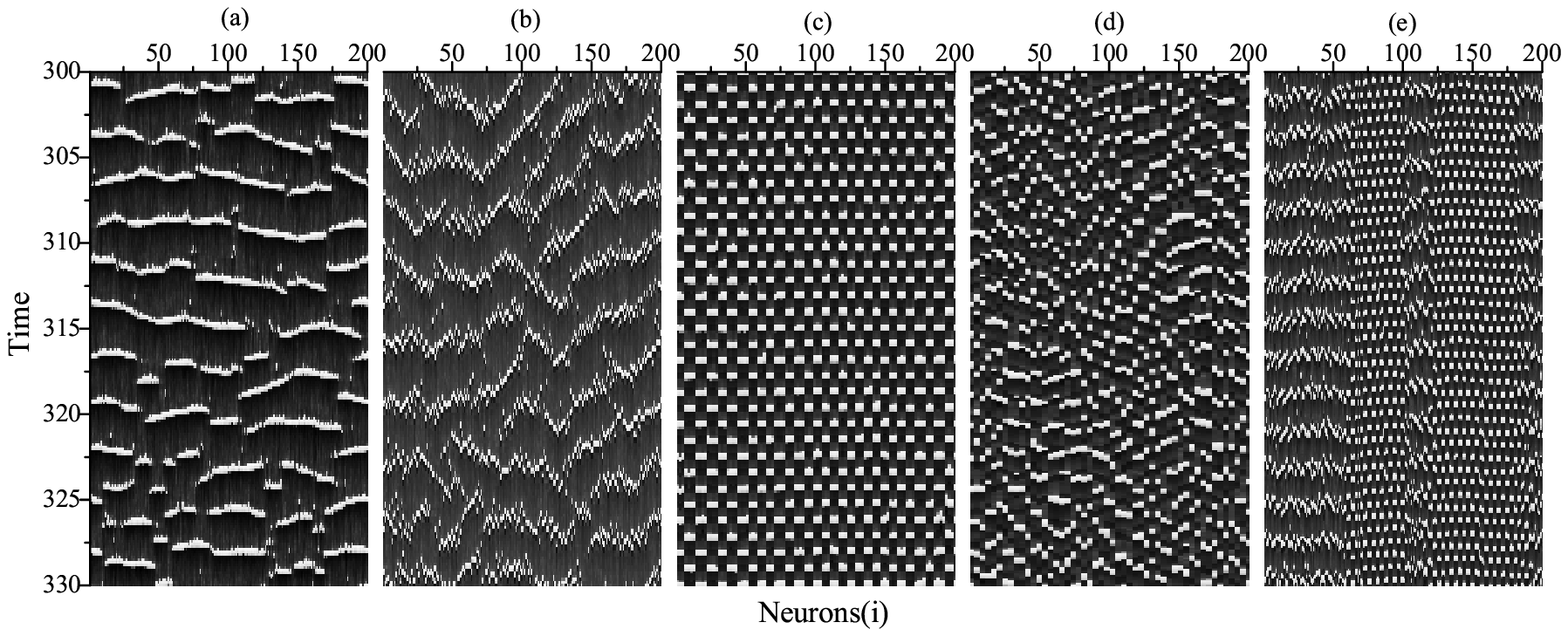}}
\caption{ Space-time plots of TW neurons with type \uppercase \expandafter {\romannumeral 2} coupling for different delay time $\tau$. From left to right, $\tau$ equals to $0, 0.3, 1.0, 1.8$ and $3.0$, respectively. Other parameters are the same as in Fig.\ref{Pattern-type1}.
\label{Pattern-type2}}
\end{figure}

In what follows, the effects of  these two types of coupling on temporal coherence and spatial synchronization in small-world networks consisting of noisy Terman-Wang (TW) excitable neurons are presented. Since we are mainly interested in the effects of delayed coupling, here we start from a regular network with periodic boundary condition consisting of $N=200$ TW neurons, each having $K=8$ nearest neighbors, and the coupling strength is fixed at $\epsilon=0.1$ throughout this paper. Results shown in Fig.\ref{Pattern-type1} and Fig.\ref{Pattern-type2} illustrate the spatiotemporal dynamics of neurons with type \uppercase \expandafter {\romannumeral 1} and type \uppercase \expandafter {\romannumeral 2} coupling, respectively. In both plots, from left to right the delay time $\tau$ equals to $0, 0.3, 1.0, 1.8,$ and $3.0$. Clearly, distinct spatiotemporal patterns can be observed with different types of coupling. In Fig.\ref{Pattern-type1}, initially, in the absence of time delay [see panel (a)] the snapshot is rather turbulent, both in space and time. With increasing $\tau$, say, $\tau=1.0$, a considerable enhancement of regularity can be observed. When the delay time is further increased, e.g., to $1.8$, as shown in the fourth panel, the system reaches a strikingly ordered state, where all the neurons are almost synchronized in space and periodic in time. However, if we further increase $\tau$, the ordered state begins to be deteriorated (e.g., $\tau=3.0$). These observations thus demonstrate that moderate time delays can enhance both temporal coherence and spatial synchronization of coupled TW neurons with type \uppercase \expandafter {\romannumeral 1} coupling . Furthermore, presented results indicate typical resonant phenomenon, i.e., for an optimal delay time, the spatiotemporal regularity of the system can reach a clear-cut maximum level. We note here that such phenomenon has already been reported in our previous paper \cite{CHA10043140}, and we present it here for clear-cut comparison with the type \uppercase \expandafter {\romannumeral 2} coupling and self-consistency of the present paper.

\begin{figure}
\centerline{\includegraphics*[width=1.0\columnwidth]{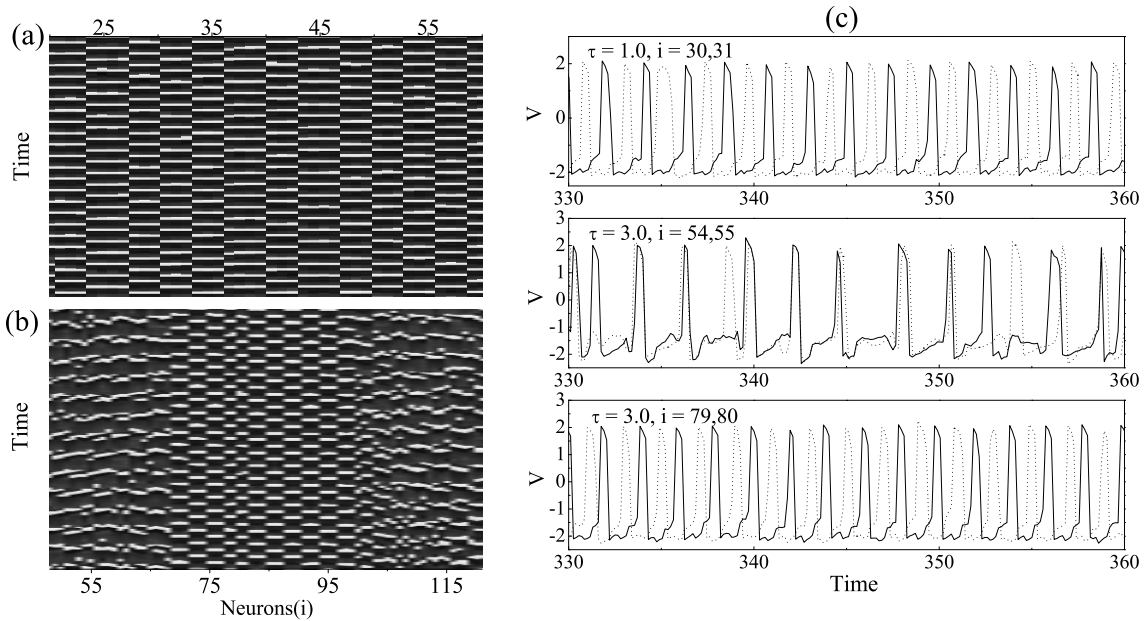}}
\caption{ (a) An inset of Fig.\ref{Pattern-type2}c, enabling a clearer demonstration of the clustering anti-phase synchronization. (b) An inset of Fig.\ref{Pattern-type2}e, displaying explicit clustered chimera states. (c) Time series of membrane potential of two nearest neurons. From top to down,  the APS state, spatial incoherence part and anti-phase synchronization part of clustered chimera states are exhibited.
\label{Type2-inset}}
\end{figure}

The situation in Fig.\ref{Pattern-type2} is totally different: we can never derive a similar spatiotemporal ordered state, but non-trivial synchronization transitions phenomena induced by time delays in type \uppercase \expandafter {\romannumeral 2} coupling can be found. When $\tau=0$, the plot is the same as Fig.\ref{Pattern-type1}(a). For non-zero yet short delays (e.g., $\tau=0.3$), zigzag fronts of excitations appear, as shown in Fig.\ref{Pattern-type2}(b). In Fig.\ref{Pattern-type2}(c), however, alternative layer waves are present where excitatory spikes occur alternatively among nearby clusters in space as the temporal dynamics evolves. Hence, this phenomenon can be termed appropriately as a clustering anti-phase synchronization (APS) transition induced by an moderate time delay, which is quite distinct from the spatiotemporal pattern in Fig.\ref{Pattern-type1}(c). As $\tau$ is further increased to $\tau=1.8$, the APS is heavily impaired. Finally, when $\tau=3.0$, intriguingly, clustered chimera states \cite{PRL08144102} which have spatially distributed anti-phase coherence separated by incoherence can be observed [see panel (e)], and this interesting phenomenon can not be found in type \uppercase \expandafter {\romannumeral 1} coupling. For clearer illustration, we plot the local enlargements as well as the time series of two nearby neurons for $\tau=1.0$ and $\tau=3.0$ in Fig.\ref{Type2-inset}, where the APS state and clustered chimera states can be shown more explicitly. Finally, the synchronization transitions displayed in Fig.\ref{Pattern-type2} could be explained by the mechanism that this type of delayed coupling can introduce phase slips, and hence zigzag fronts and alternative layer waves even chimera states can appear, thus supplementing the purely noise-induced excitations.

To quantitatively characterize the spatiotemporal dynamics of the neuronal systems, we introduce the coefficient of variance (CV) of the inter-spike intervals (ISI) and the standard deviation factor $\sigma$ to measure temporal coherence and spatial synchronization, separately \cite{PRL97000775, CPC05001402, CPC06000579}. CV is defined as
\begin{align}
\lambda_i &= \f{\la T_i \ra_t}{\sqrt{\la T_i^2 \ra_t - \la T_i \ra_t^2 }}
\label{lamda}
\end{align}

\noindent where $\la \cdot \ra_t$ denotes averaging over time, $T_i$ is the ISI of neuron $i$ . By further
averaging $\lambda_i$ over different neurons, we get $\lambda=\sum_i \lambda_i/N$ as the CV of the coupled neuronal network. Obviously a larger $\lambda$ means better periodicity in time. The standard deviation factor is defined as $\sigma=\la \sigma (t) \ra_t$, where
\begin{align}
\sigma(t) &= \sqrt{\f{(\sum_i x^2_i(t))/N-(\sum_i x_i(t)/N)^2}{N-1}}
\label{sigma}
\end{align}

\noindent Clearly a smaller $\sigma$ means better synchronization in space. Final results shown below are averaged over $50$ independent runs for each set of parameter values to warrant appropriate statistical accuracy with respect to the network generation and numerical simulations.

\begin{figure}
\centerline{\includegraphics*[width=1.0\columnwidth]{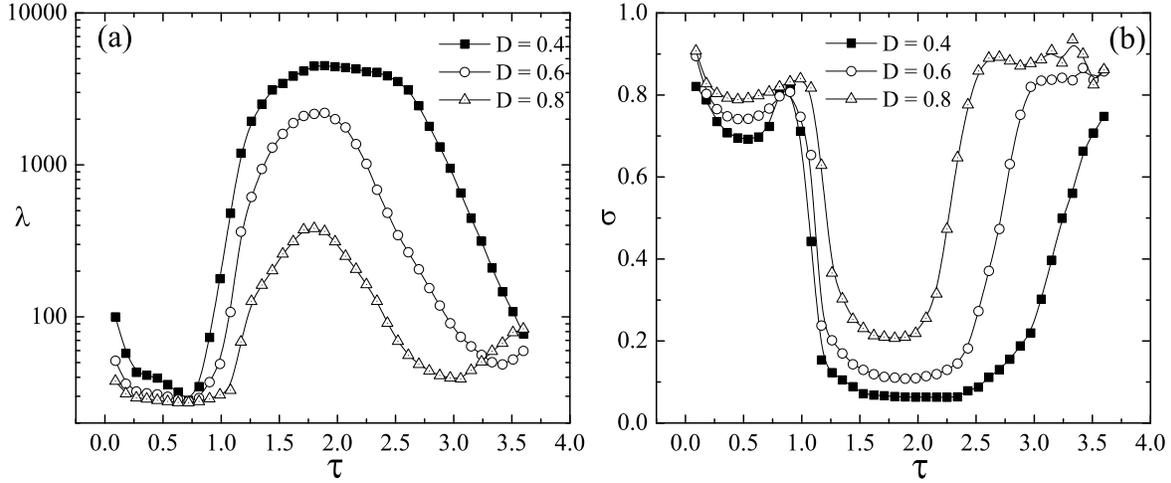}}
\caption{ (a) Dependence of $\lambda$ on the delay time $\tau$ for different noise intensity $D$. (b) Dependence of $\sigma$ on $\tau$ for different $D$. The coupling type is type \uppercase \expandafter {\romannumeral 1}. Other parameters are the same as in Fig.\ref{Pattern-type1}
\label{curve-type1}}
\end{figure}

\begin{figure}
\centerline{\includegraphics*[width=1.0\columnwidth]{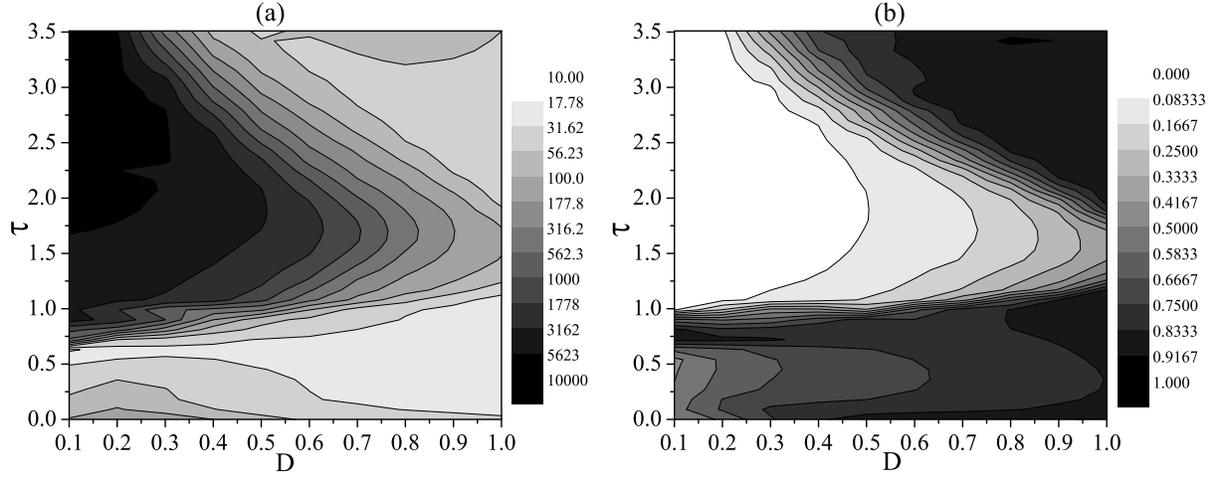}}
\caption{ (a) Contour plot of $\lambda$ in dependence on the delay time $\tau$ and the noise intensity $D$. (b) Contour plot of $\sigma$ in dependence on $\tau$ and $D$. The coupling type is type \uppercase \expandafter {\romannumeral 1}. Delay-induced stochastic resonance is clearly visible. Other parameters are the same as in Fig.\ref{curve-type1}
\label{contour-plot1}}
\end{figure}

In Fig.\ref{curve-type1}, we have plotted $\lambda$ and $\sigma$ versus delay time $\tau$ for different values of noise intensity $D$ for type \uppercase \expandafter {\romannumeral 1} delay. In accordance with the visual inspection of Fig.\ref{Pattern-type1}, delay-induced resonances in $\lambda$ and $\sigma$ depending upon the increase of $\tau$ are observed, where clear-cut peaks and valleys occur at the optimal delay time $\tau_{opt1} \simeq 1.8$, corresponding to the most ordered spatiotemporal state ploted in Fig.\ref{Pattern-type1}(d). Following the common terminology, these are termed as delay-induced stochastic resonance on neuronal networks. Moreover, it is clear that as $D$ increases, the peak value in the $\lambda \sim \tau$ curve decreases monotonically and the valley in the $\sigma \sim \tau$ curve gets higher, respectively, which means increasing the noise intensity $D$ can impair the resonance phenomenon, but the particular location of the $\tau_{opt1}$ are robust to the change of $D$. To make an overall inspection, the dependence of $\lambda$ and $\sigma$ on both delay time $\tau$ and noise intensity $D$ is shown in Fig.\ref{contour-plot1}. It is evident that the results presented in Fig.\ref{curve-type1} keep robust in a considerable range of $\tau$ and $D$.

\begin{figure}
\centerline{\includegraphics*[width=1.0\columnwidth]{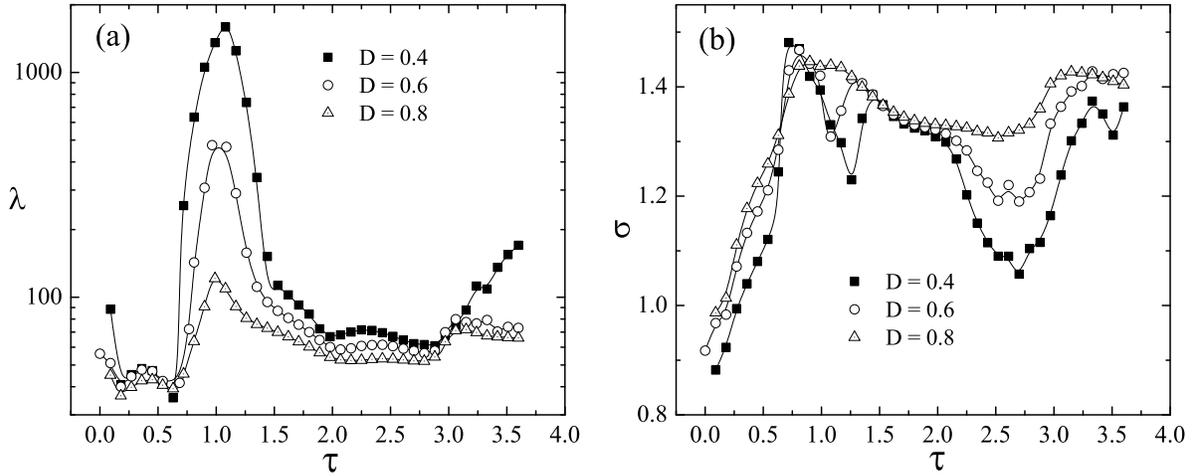}}
\caption{ (a) Dependence of $\lambda$ on the delay time $\tau$ for different noise intensity $D$. (b) Dependence of $\sigma$ on $\tau$ for different $D$. The coupling type is type \uppercase \expandafter {\romannumeral 2}. Other parameters are the same as in Fig.\ref{Pattern-type2}
\label{curve-type2}}
\end{figure}

Next, we quantitatively study the impact of type \uppercase \expandafter {\romannumeral 2} delayed coupling on the spatiotemporal dynamics of the neuronal networks. Numerical results in Fig.\ref{curve-type2} illustrate the dependence of $\lambda$ and $\sigma$ on the delay time $\tau$ for different noise intensity $D$. As visually interpreted by space-time plots in Fig.\ref{Pattern-type2}, a spatiotemporal ordered state can not be achieved, but non-trivial synchronization transitions induced by time delay appear. For a given noise intensity $D$, we can see that as $\tau$ is increased (e.g.,$\tau=0.3$), $\sigma$ increases sharply, corresponding to the appearance of zigzag fronts which destroys synchronization. With further increasing delay, $\lambda$ and $\sigma$ pass through a peak at about $\tau=0.9$, indicating clustering APS state which is periodic in time but poor in synchronization. When $\tau$ increases again, e.g., to $\tau=1.8$, $\lambda$ and $\sigma$ decrease clearly, corresponding to the deterioration of the APS state [See Fig.\ref{Pattern-type2}(d)]. However, when $\tau$ is larger than $\tau=2.7$, $\lambda$ and $\sigma$ begin to increase again, which shows, in accordance with the visual inspection of Fig.\ref{Pattern-type2}(e), the emergence of clustered chimera states that have spatially distributed anti-phase coherence separated by incoherence. We can easily understand that the increasing of $\lambda$ and $\sigma$ comes from the anti-phase coherence part of the chimera states. In addition, the effect of the noise intensity $D$ is similar to the results shown in Fig.\ref{curve-type1}, that is to say, increasing the noise intensity $D$ can dent the synchronization transitions phenomena but keeps the qualitative behaviors unchanged.

\begin{figure}
\centerline{\includegraphics*[width=1.0\columnwidth]{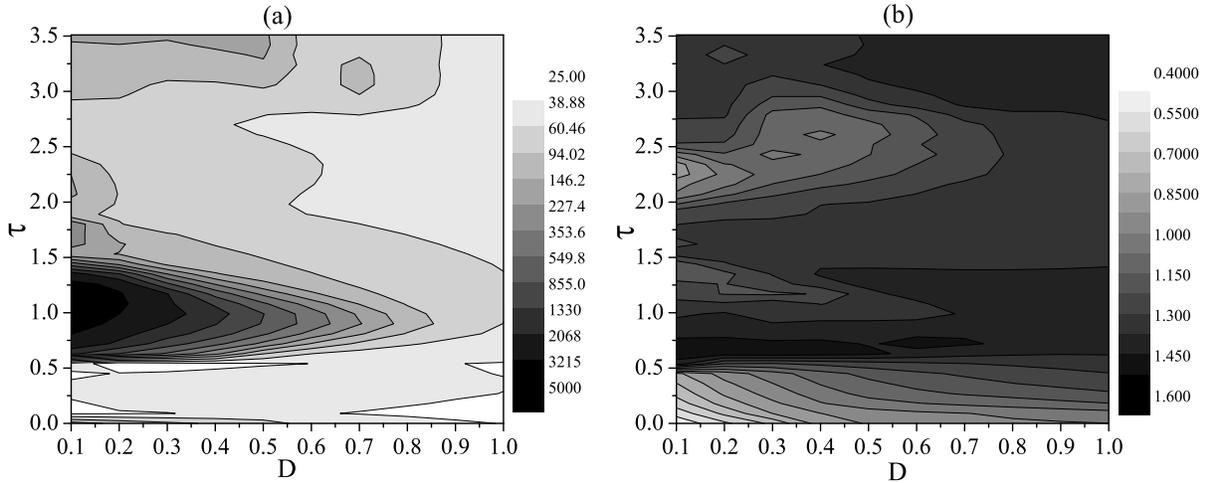}}
\caption{ (a) Contour plot of $\lambda$ in dependence on the delay time $\tau$ and the noise intensity $D$. (b) Contour plot of $\sigma$ in dependence on $\tau$ and $D$. The coupling type is type \uppercase \expandafter {\romannumeral 2}. Other parameters are the same as in Fig.\ref{curve-type2}
\label{contour-plot2}}
\end{figure}

In Fig.\ref{contour-plot2}, we display the contour plots of the dependence of $\lambda$ and $\sigma$ on both delay time $\tau$ and noise intensity $D$ for type \uppercase \expandafter {\romannumeral 2} delay. It is evident the delay-induced synchronization transitions are robust in a large area of $\tau$ and $D$. Furthermore, the particular location of the $\tau_{opt2}$ where the peak of $\lambda$ appears, in accordance with the anti-phase synchronization state, stays almost the same as $D$ is varied, see Fig.\ref{contour-plot2}(a). The value of $\tau_{opt2}$ is about $0.9$, which is just half of the optimal delay time $\tau_{opt1}$ where the spatiotemporal ordered state emerges as shown in Fig.\ref{contour-plot1}. We will return to this nontrivial point later and give some qualitative analysis in the discussion section.

\begin{figure}
\centerline{\includegraphics*[width=1.0\columnwidth]{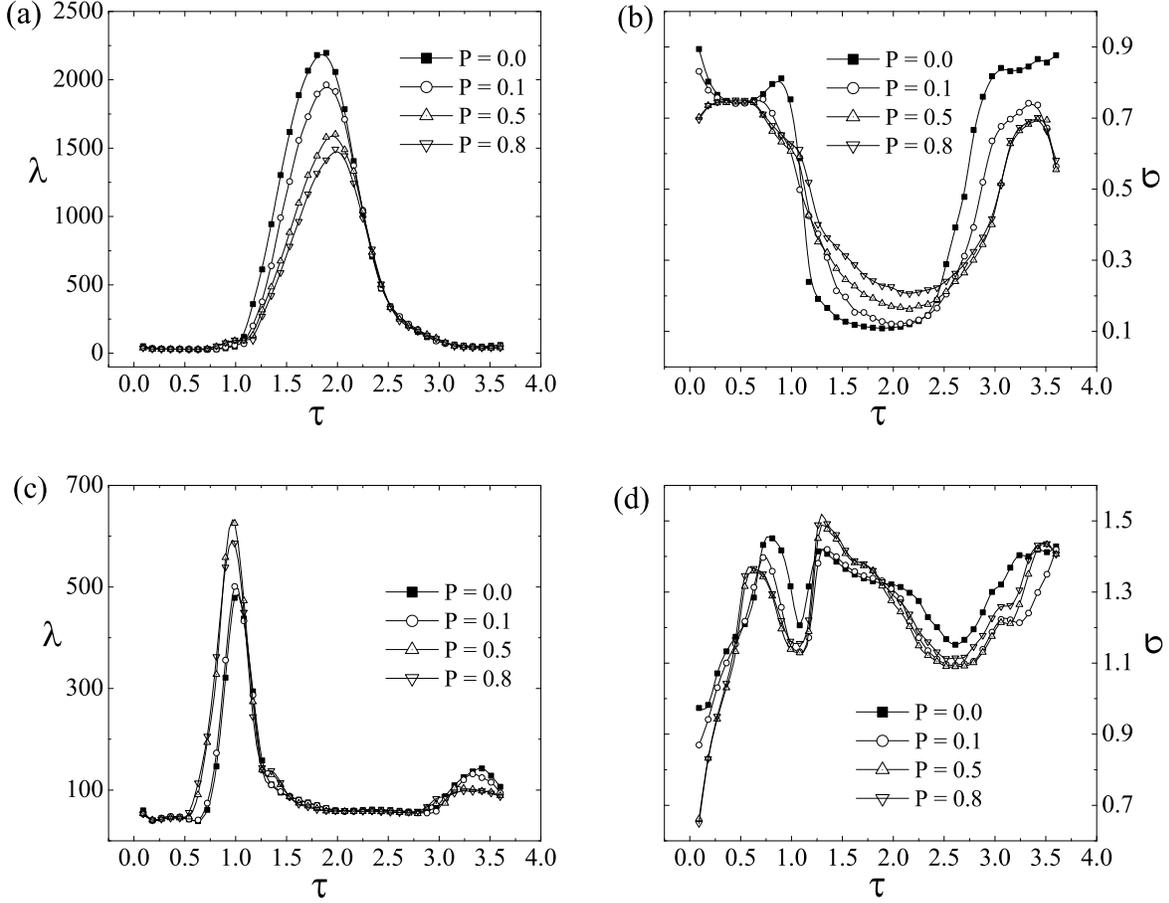}}
\caption{ Dependence of $\lambda$ and $\sigma$ on the delay time $\tau$ for different rewiring probability $p$. (a), (b) Type \uppercase \expandafter {\romannumeral 1} delayed coupling. (c), (d) Type \uppercase \expandafter {\romannumeral 2} delayed coupling. Other parameter values are $N=200$, $K=8$, and $D=0.6$.
\label{change-p}}
\end{figure}

Actually, real neuronal networks often have complex topology. In recent years, neuronal dynamics on complex networks, e.g., small-world (SW) ones \cite{NTR98000440}, has actually become a focal research topic in theoretical neuroscience \cite{PRL00002758, PRL06238103, PRL07108101, CPC05001402, CPC06000579, CSF07000280}, and network topology could play vital role in neuronal synchronization or coding dynamics. In this work, we also address such issues by performing similar studies on SW networks. We generate SW networks following the Watts-Strogatz scheme by rewiring the edges in a regular network with probability $p$. The network changes from being regular to totally random with $p$ from 0 to 1, but keeping the number of total links unchanged. We plot the dependence of spatiotemporal dynamics upon the delay time $\tau$ for different rewiring probability $p$ with type \uppercase \expandafter {\romannumeral 1} and type \uppercase \expandafter {\romannumeral 2} coupling in Fig.\ref{change-p}. It is noteworthy that the qualitative results achieved on regular networks above, i.e., the constructive roles of the two types of delayed coupling, are robust against the rewiring probability $p$ of the small-world network, but with some tiny quantitative differences. For type \uppercase \expandafter {\romannumeral 1} coupling [see panel (a), (b)], as the network becomes more and more random ($p$ increases), just similar to the effects of increasing noise intensity $D$, the peak value in the $\lambda \sim \tau$ curve decreases monotonically and the valley in the $\sigma \sim \tau$ curve gets higher, respectively, which means increasing the rewiring probability $p$ can deteriorate the resonance phenomenon. Moreover, the optimal delay time $\tau_{opt1}$ where clear-cut peaks and valleys appear almost retains unchanged ($\tau_{opt1} \simeq 1.8$), indicating that $\tau_{opt1}$ is not sensitive to the rewiring probability $p$. Whereas for type \uppercase \expandafter {\romannumeral 2} coupling [see Fig.\ref{change-p}(c) and (d)], the peak value in the $\lambda \sim \tau$ curve increases non-monotonically as $p$ is increased, and just like $\tau_{opt1}$, the value of $\tau_{opt2}$ always fixes at about $0.9$.

\section{Discussion and Conclusion\label{Section4}}

\begin{figure}
\centerline{\includegraphics*[width=1.0\columnwidth]{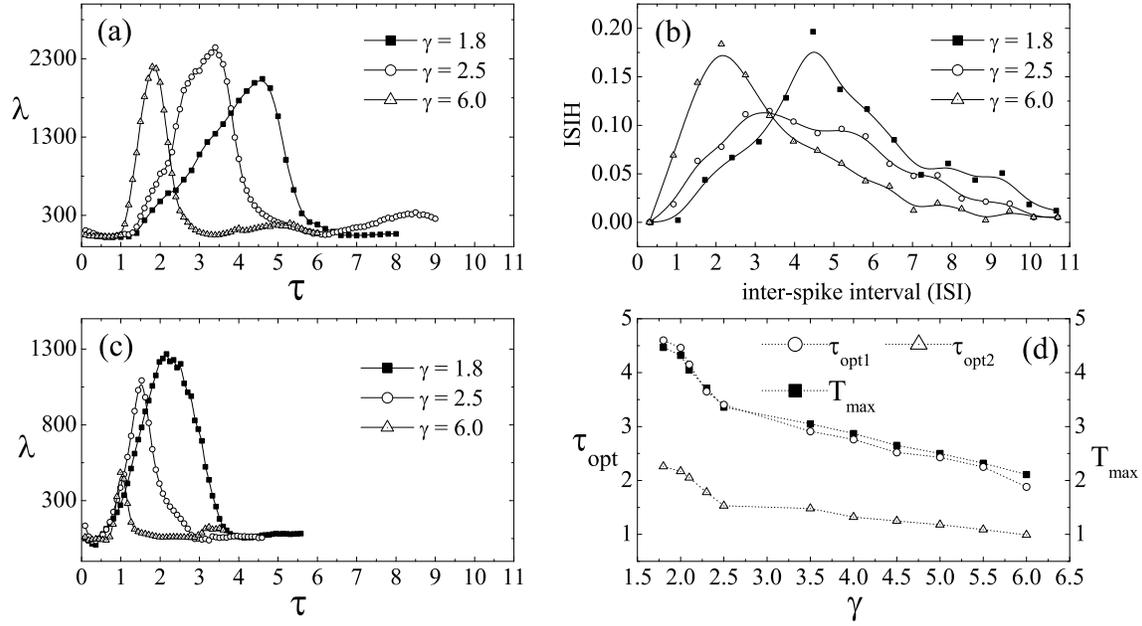}}
\caption{ (a), (c) Dependence of $\lambda$ on the delay time $\tau$ for different model parameter $\gamma$ with type \uppercase \expandafter {\romannumeral 1} and type \uppercase \expandafter {\romannumeral 2} delayed coupling, respectively. (b) Normalized ISIH of coupled noisy TW neuronal networks without time delay for different $\gamma$. (d) Dependence of $\tau_{opt1}$,$\tau_{opt2}$ and $T_{max}$ on the parameter $\gamma$. Other parameters are the same as in Fig.\ref{change-p}
\label{scale-relation}}
\end{figure}

As stated before, both the optimal delay time $\tau_{opt1}$ and $\tau_{opt2}$ are robust to the changing of the noise intensity $D$ and the rewiring probability $p$, and furthermore, $\tau_{opt2}$ is almost half of $\tau_{opt1}$. We are thus wondering what is the underline mechanism of such an interesting phenomenon, and finally find that it is relevant with some intrinsic time scale of the systems. In Fig.\ref{scale-relation}, we show the relationship of $\tau_{opt1}$ and $\tau_{opt2}$ to the intrinsic time scale of the systems. Fig.\ref{scale-relation}(a) and (c) depict the dependence of $\lambda$ on the delay time $\tau$ for the model parameter $\gamma=1.8, 2.5$ and $6.0$ with type \uppercase \expandafter {\romannumeral 1} and type \uppercase \expandafter {\romannumeral 2} coupling, respectively. We can see clearly the optimal delay time $\tau_{opt1}$ and $\tau_{opt2}$ decrease as $\gamma$ increases, more interestingly, for a equal $\gamma$ the value of $\tau_{opt2}$ is just half of $\tau_{opt1}$. Accordingly, we have calculated the normalized inter-spike interval histograms (ISIHs) of coupled neuronal networks without time delay to investigate the inherent spike period of the neuronal networks, as shown in Fig.\ref{scale-relation}(b). Obviously, the peak position of the ISIH matches with $\tau_{opt1}$ quite well, and is just twice the value of $\tau_{opt2}$. In Fig.\ref{scale-relation}(d), we give more quantitative illustration, where $T_{max}$ is the peak position of the ISIH, meaning the inherent time scale of the neuronal systems. It can be observed that as $\gamma$ changes from $1.8$ to $6.0$, the optimal delay time $\tau_{opt1}$ where the spatiotemporal ordered states emerges always equals to $T_{max}$, which represents a kind of locking between the delay time and inherent spiking period of the neuronal network under the effects of noise. Whereas for type \uppercase \expandafter {\romannumeral 2} coupling, $\tau_{opt2}$ always keeps the value half of $T_{max}$, the reason may be that type \uppercase \expandafter {\romannumeral 2} delayed coupling can pull adjacent neurons into anti-phase synchronization, the optimal delay time $\tau_{opt2}$ warranting the best spike regularity is not equal to one spiking period. Thus, it is exactly the half of the inherent spiking period of the neuronal network, where the phase locking between antiphased spikes occurs. In addition, we have also tried some other relaxation oscillator models which can describe dynamics of neurons, such as FHN model, and obtain similar qualitatively results.

In summary, We have investigated temporal coherence and spatial synchronization on small-world networks consisting of noisy Terman-Wang excitable neurons in dependence on two types of time-delayed coupling. For type \uppercase \expandafter {\romannumeral 1} coupling, we show that time delay can dramatically enhance temporal coherence and spatial synchrony of the noise-induced spike trains, and if the delay time is tuned to nearly match the intrinsic spike period of the neuronal network, the system dynamics reaches a most ordered state, which is periodic in time and nearly synchronized in space, demonstrating an interesting type of resonance phenomenon with delay. For type \uppercase \expandafter {\romannumeral 2} coupling, however, a similar spatiotemporal ordered state never appears, but as the delay time is increased, the neurons exhibit synchronization transition from zigzag fronts of excitations to dynamic clustering anti-phase synchronization, and further to clustered chimera states that have spatially distributed anti-phase coherence separated by incoherence. Furthermore, we also show that these findings are robust to the changing of the noise intensity $D$ and the rewiring probability $p$. Finally, qualitative analysis is given to illustrate the numerical results. Since time delays are inevitable in real neuronal systems, we hope that our results will be helpful for further understanding the roles of delay in neuron firing on realistic neuronal networks.

\section*{Acknowledgement}
This work is supported by National Science Foundation of China (20933006, 20873130) and
 the Fundamental Research Funds for the Central Universities.

%

\bibliography{mns}
\bibliographystyle{apsrev}

\end{document}